# Theta and/or alpha? Neural oscillational substrates for dynamic inter-brain synchrony during mother-child cooperation

Jiayang Xu[1], Yamin Li[1,2], Ruxin Su[3], Saishuang Wu[4], Chengcheng Wu[1], Haiwa Wang[4], Qi Zhu[4], Yue Fang[5], Fan Jiang[4], Shanbao Tong[1], Yunting Zhang[6, *], Xiaoli Guo[1, *]

[1] School of Biomedical Engineering, Shanghai Jiao Tong University, Shanghai, China.
[2] Department of Computer Science, Vanderbilt University, Nashville, TN, USA.
[3] Shanghai Mental Health Center, Shanghai Jiao Tong University School of Medicine, Shanghai, China.
[4] Department of Developmental and Behavioral Pediatrics, National Children's Medical Center, Shanghai Children's Medical Center, affiliated to School of Medicine Shanghai Jiao Tong University, Shanghai, China
[5] China Welfare Institute Nursery, Shanghai, China.
[6] Child Health Advocacy Institute, National Children's Medical Center, Shanghai Children's Medical Center, affiliated to School of Medicine Shanghai Jiao Tong University, Shanghai, China.

**\* Corresponding authors:**
Dr. Xiaoli Guo at School of Biomedical Engineering, Shanghai Jiao Tong University, Shanghai, China, e-mail: meagle@sjtu.edu.cn.
Dr. Yunting Zhang at Child Health Advocacy Institute, National Children's Medical Center, Shanghai Children's Medical Center, affiliated to School of Medicine Shanghai Jiao Tong University, Shanghai, China, e-mail: edwinazhang@shsmu.edu.cn.

**Abstract:**

Mother-child interaction is a highly dynamic process neurally characterized by inter-brain synchrony (IBS) at $\theta$ and/or $\alpha$ rhythms. However, their establishment, dynamic changes, and roles in mother-child interactions remain unknown. Through dynamic analysis of dual-EEG from 40 mother-child dyads during turn-taking cooperation, we uncover that $\theta$-IBS and $\alpha$-IBS alternated with interactive behaviors, with EEG frequency-shift as a prerequisite for IBS transitions. When mothers attempt to track their children's attention and/or predict their intentions, they will adjust their EEG frequencies to align with their children's $\theta$ oscillations, leading to a higher occurrence of the $\theta$-IBS state. Conversely, the $\alpha$-IBS state, accompanied by the EEG frequency-shift to the $\alpha$ range, is more prominent during mother-led interactions. Further exploratory analysis reveals greater presence and stability of the $\theta$-IBS state during cooperative than non-cooperative conditions, particularly in dyads with stronger emotional attachments and more frequent interactions in their daily lives. Our findings shed light on the neural oscillational substrates underlying the IBS dynamics during mother-child interactions.

# 1. Introduction

Mother-child synchrony, involving dynamic and reciprocal adjustments in behavior and neurophysiological processes between each other (Feldman, 2012; George, 1996), is regarded as a key aspect of mother-child interactions (Leclère et al., 2014). Of the various forms of synchrony, neural synchrony between two brains has been considered a biomarker for shared emotional engagement, fostering an environment of better learning and cognitive development for children (Atzil and Gendron, 2017; Nguyen et al., 2020; Wass et al., 2020).

With the advent of hyperscanning techniques (Czeszumski et al., 2020; Lindenberger et al., 2009; Marriott Haresign et al., 2022; Wang et al., 2018, 2020), researchers are now capable of measuring inter-brain synchrony (IBS) during mother-child interactions by simultaneously monitoring two brains (Levy et al., 2017; Miller et al., 2019; Nguyen et al., 2021, 2020; Reindl et al., 2018; Zhao et al., 2021). As an example, IBS derived from dual-EEG can characterize phase entrainment and rhythmic co-fluctuations between two brains on a subsecond scale (Marriott Haresign et al., 2022), acting as a proxy measure for the dynamic interaction process. Using dual-EEG recording and analysis, researchers have found synchrony between the brain oscillations of mother and child. This synchrony is higher on average when compared to stranger-child dyads (Endevelt-Shapira et al., 2021), in more engaged interactions (like cooperation) (Li et al., 2024), and in the presence of specific communication signals such as maternal body odors (Endevelt-Shapira et al., 2021) and positive emotion (Santamaria et al., 2020). However, these studies only investigated the IBS averaged across the whole interaction condition or even the whole interaction process (also known as "static analysis"), thus failing to reveal any dynamic properties of IBS. Nevertheless, the mixed findings in IBS at different EEG rhythms suggest that multi-rhythmic co-fluctuations might occur between children's and mothers' brain oscillations. These co-fluctuations may inter-switch during interactions, potentially being an inherent feature of IBS dynamics.

More specifically, the $\theta$ and $\alpha$ rhythms have been consistently shown to be responsible for mother-child inter-brain communication (Marriott Haresign et al., 2022; Turk et al., 2022; Wass et al., 2018). However, synchrony of $\theta$, $\alpha$ or both rhythms between mother and child were inconclusively reported in previous studies in static analysis (Atilla et al., 2023; Endevelt-Shapira et al., 2021; Endevelt-Shapira and Feldman, 2023; Leong et al., 2017; Santamaria et al., 2020). Leong et al. found that direct gaze at infants increased bidirectional inter-brain connectivity in both $\theta$ and $\alpha$ bands (Leong et al., 2017). Santamaria et al. found that maternal emotions would influence the mother-infant inter-brain network in the α band (Santamaria et al., 2020). Endevelt-Shapira et al. found that θ-band IBS in mother-infant was higher than that in stranger-infant, which was linked to maternal body odors, sensitivity, and intrusiveness (Endevelt-Shapira et al., 2021; Endevelt-Shapira and Feldman, 2023). Atilla et al. reported higher α-IBS when a child was observing the mother performing the task compared to when the mother was observing her child (Atilla et al., 2023). Considering the inconsistent findings in IBS rhythms so far, we speculate that during the highly dynamic mother-child interactions, there are likely two distinct IBS states in which two brains synchronize at θ and α rhythms, respectively. However, these two IBS states were compounded when averaged across the interaction process, which may contribute to the variability in previous static studies. This speculation was also echoed by our recent study which observed a transition between θ-IBS and α-IBS along with role alternation during mother-child cooperation (Li et al., 2024). This exploratory study suggested an association of mother-child IBS dynamics with EEG rhythmic change; however, calculating and averaging IBS by role was essentially limited to static analysis and was insufficient to uncover the highly dynamic IBS states during the back-and-forth interactions.

Learning how IBS is established at θ and α rhythms is another key to understanding the mother-child inter-brain communication during reciprocal interactions. The θ rhythm is crucial for infants and preschoolers in cognitive functioning, anticipatory and sustained attention, and encoding (Meyer et al., 2019; Orekhova et al., 2006; Saby and Marshall, 2012; Sperdin et al., 2018; Wass et al., 2018); while for adults, the prime oscillations related to attention and cognition are within α band (Bazanova and Vernon, 2014; Marshall et al., 2002; Wass et al., 2018). So how do two brains with different prime frequencies of oscillations synchronize efficiently during mother-child interactions? Some researchers have proposed a hypothesis of cross-frequency synchrony (Noreika et al., 2020); however, a frequency-shift phenomenon found by Wass et al. prompts us with another solution through within-frequency synchrony. Wass et al. observed that during mother-infant joint play, the attention-related EEG frequency of mothers was down-shifted from their α band to infants' attention-related θ range (Wass et al., 2018). We speculate this mother-to-child frequency-shift to be a prerequisite for establishing IBS at θ band. Likewise, children may up-shift their EEG rhythm to the α range, so as to align with their mothers' EEG and achieve an α-IBS state. Furthermore, the direction of EEG frequency-shifts can provide further insights into the roles of θ-IBS and α-IBS in mother-child inter-brain communication. That is, the mother-to-child frequency-shift suggests that *θ-IBS* may play a pivotal role during child-led interactions when mothers were attempting to track children's attention and/or predict children's intentions. Conversely, *α-IBS* may be more critical during mother-led interactions. To capture the EEG frequency-shifts along with IBS state transitions and the dynamic fluctuations of IBS along with child-led and mother-led behaviors, a simultaneous dynamic analysis of intra-brain EEG power, inter-brain synchrony, and behavior coding is needed.

Therefore, to fill the knowledge gaps about how IBS is established and dynamically changing during mother-child interactions, we designed the following studies. First, we constructed a dual-band inter-brain network framework that comprised *θ*-band and *α*-band inter-brain subnetworks, and proposed a data-driven approach utilizing sliding windows and *k*-means clustering to capture the dynamic IBS states during mother-child interactions **(Figure 1)**. Through dynamic inter-brain network analysis of dual-EEG data from 40 mother-child dyads engaged in a turn-taking tangram puzzle-solving task, we aim to test **Hypothesis 1: during mother-child interactions, there are two distinct IBS states characterized by inter-brain synchronization within *θ* and *α* bands, respectively**. Then, in order to understand the different roles of *θ-IBS* and *α-IBS* in mother-child inter-brain communication, we analyzed the dynamic changes of IBS states along with child-led and mother-led behaviors in turn-taking cooperation, to test **Hypothesis 2: *θ-IBS* and *α-IBS* are crucial for child-led and mother-led interactions, respectively.** Third, we calculated the cross-correlation between the dynamics of EEG power and IBS states, to test **Hypothesis 3** about how the two IBS states are established: ***θ-IBS* state is achieved by mother-toward-child neural alignment with a down-shift of mothers' EEG frequencies to children's *θ* range; conversely, *α-IBS* state is achieved by child-toward-mother neural alignment, with an up-shift of children's EEG frequencies to the *α* range**. In addition, we compared dynamic IBS states between 18 mother-child dyads and 19 stranger-child dyads and between their individually-solving and cooperatively-solving conditions, to explore important dynamic characteristics linked to relationships and interactions.

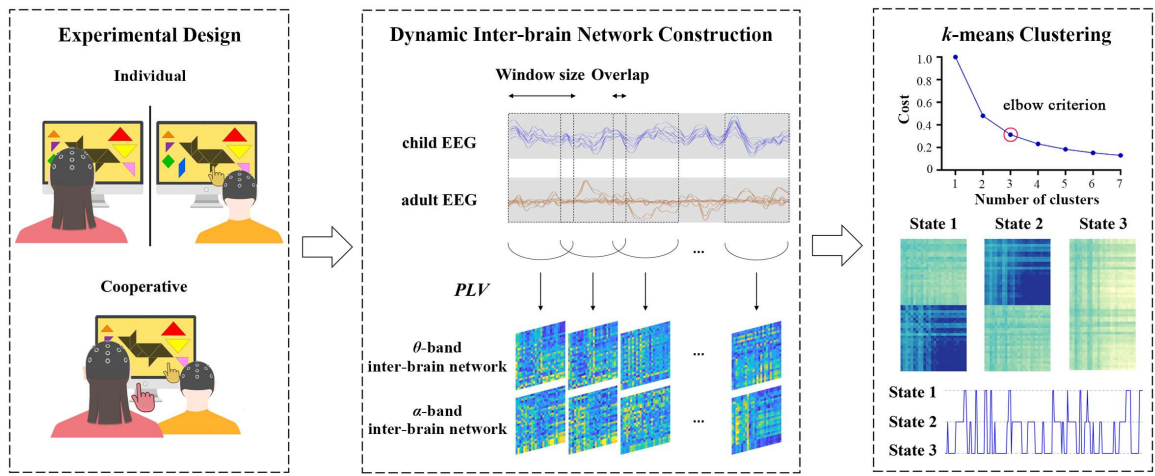

**Figure 1. The flowchart of dynamic inter-brain network analysis.** The dual-EEG data used for dynamic analysis in this study include those of mother-child dyads and stranger-child dyads when solving a tangram puzzle task individually or cooperatively. In the individual condition, one participant solved all the puzzles alone, and the other was seated in another room, watching the screen recording from the actor. In the cooperative condition, the participants were seated next to each other and solved tangram templates jointly by moving the puzzle pieces by turn. In the dynamic analysis, dual-EEG data are firstly segmented using a sliding window. Then, in each time window, a dual-band inter-brain network comprising two subnetworks in $\theta$ (3-6 Hz) and $\alpha$ (6-9 Hz) band, respectively, is constructed using phase locking value (PLV). Next, dual-band inter-brain networks from all dyads and all time windows are clustered using the $k$-means clustering method. The number of clusters $k$ is determined by the elbow criterion of cluster cost, which has shown optimal $k = 3$ either in mother-child cooperation or when combining the data from both groups and both interactive conditions. The clustering is repeated 1000 times with different initializations to escape the local minima. Finally, the final cluster centroids as well as the time series of states are obtained.

## 2. Results

**Dynamic IBS states during mother-child cooperation**

Through approaches of sliding windows and $k$-means clustering, we found three recurrent IBS states during mother-child cooperation, including one weakly connected state (*weak-IBS* state) and two strongly connected states at $\theta$ and $\alpha$ rhythms, respectively ($\theta$-*IBS* state and $\alpha$-*IBS* state) **(Figure 2A)**. The strongly-connected $\theta$-network in $\theta$-*IBS* state and $\alpha$-network in $\alpha$-*IBS* state showed a high spatial similarity (Pearson's $r = .96$) between each other **(Figure 2B)**., with strongest inter-brain connections to be established primarily between children's and mothers' central, parietal, and occipital areas **(Figure 2C)**. One-way repeated measure ANOVA of IBS strength demonstrated that *θ-IBS* state showed a significantly higher IBS at *θ* band compared with the other two states (*θ-IBS* state: 0.478 ± 0.002, *α-IBS* state: 0.388 ± 0.003, *weak-IBS* state: 0.365 ± 0.002; $F(2, 78) = 961$, $p < .001$, $\eta_p^2 = 0.961$, both post-hoc $p_{Bonf} < .001$, Fig.2D); while *α-IBS* state showed a significantly higher IBS at *α* band compared with the other two states (*α-IBS* state: 0.473 ± 0.003, *θ-IBS* state: 0.385 ± 0.002, *weak-IBS* state: 0.363 ± 0.001; $F(2, 78) = 1428$, $p < .001$, $\eta_p^2 = 0.973$, both post-hoc $p_{Bonf} < .001$, **Figure 2D**).

Among the three dynamic IBS states, the occurrence (i.e., fraction time) of *weak-IBS* state reached an average percentage of 0.485 ± 0.008, which was significantly higher than the other two strongly connected states (*θ-IBS* state: 0.266 ± 0.007, *α-IBS* state: 0.249 ± 0.008, $F(2, 78) = 221$, $p < .001$, $\eta_p^2 = 0.850$, both post-hoc $p_{Bonf} < .001$, **Figure 2D**). Similarly, the dwell time of *weak-IBS* state (589.22 ± 12.51 ms) was significantly longer than *θ-IBS* state (394.02 ± 9.39 ms) and *α-IBS* state (394.02 ± 9.39 ms) ($F(2, 78) = 112$, $p < .001$, $\eta_p^2 = 0.741$, both post-hoc $p_{Bonf} < .001$, **Figure 2D**). However, no statistically significant difference in the fraction time or dwell time was observed between *θ-IBS* state and *α-IBS* state (both post-hoc $p_{Bonf} > .5$). The laterality index (LI) between the occurrence of *θ-IBS* state and *α-IBS* state was 0.017 ± 0.028.

Although *θ-IBS* state and *α-IBS* state were overall balanced in occurrence across the whole cooperation (whole-round LI compared with zero: $t(31) = 0.590$, $p = .559$, Cohen's d = 0.104), their proportions were found dynamically changed with the turn-taking cooperative behaviors in the procedure (**Figure 2E**). When mothers were observing their children solve a puzzle (i.e., child-led puzzle-solving phase), *θ-IBS* state was more prevalent than *α-IBS* state, with a positive LI (0.112 ± 0.046) in this phase (compared with whole-round LI: $t(31) = 2.41$, $p_{FDR} = .044$, Cohen's d = 0.426, **Figure 2F**). Conversely, a negative LI (-0.081±0.038) was found when children were observing their mothers move the right hand to pick up a puzzle piece in preparation for puzzle-solving (i.e., mother-led preparation phase) (compared with whole-round LI: $t(31) = -2.83$, $p_{FDR} = .032$, Cohen's d = -0.501, **Figure 2F**), indicating a higher occurrence of *α-IBS* state than *θ-IBS* state in this phase. However, the other two phases did not show any significant changes in LI compared with whole-round LI (both $p_{FDR} > .62$, Cohen's |d| < 0.131). Moreover, the LI in the child-led puzzle-solving phase was significantly and positively correlated with the score of secure attachment (SA) (Pearson's $r = .451$, 95% CI: [.122, .691], $p = .010$, **Figure 2G**).

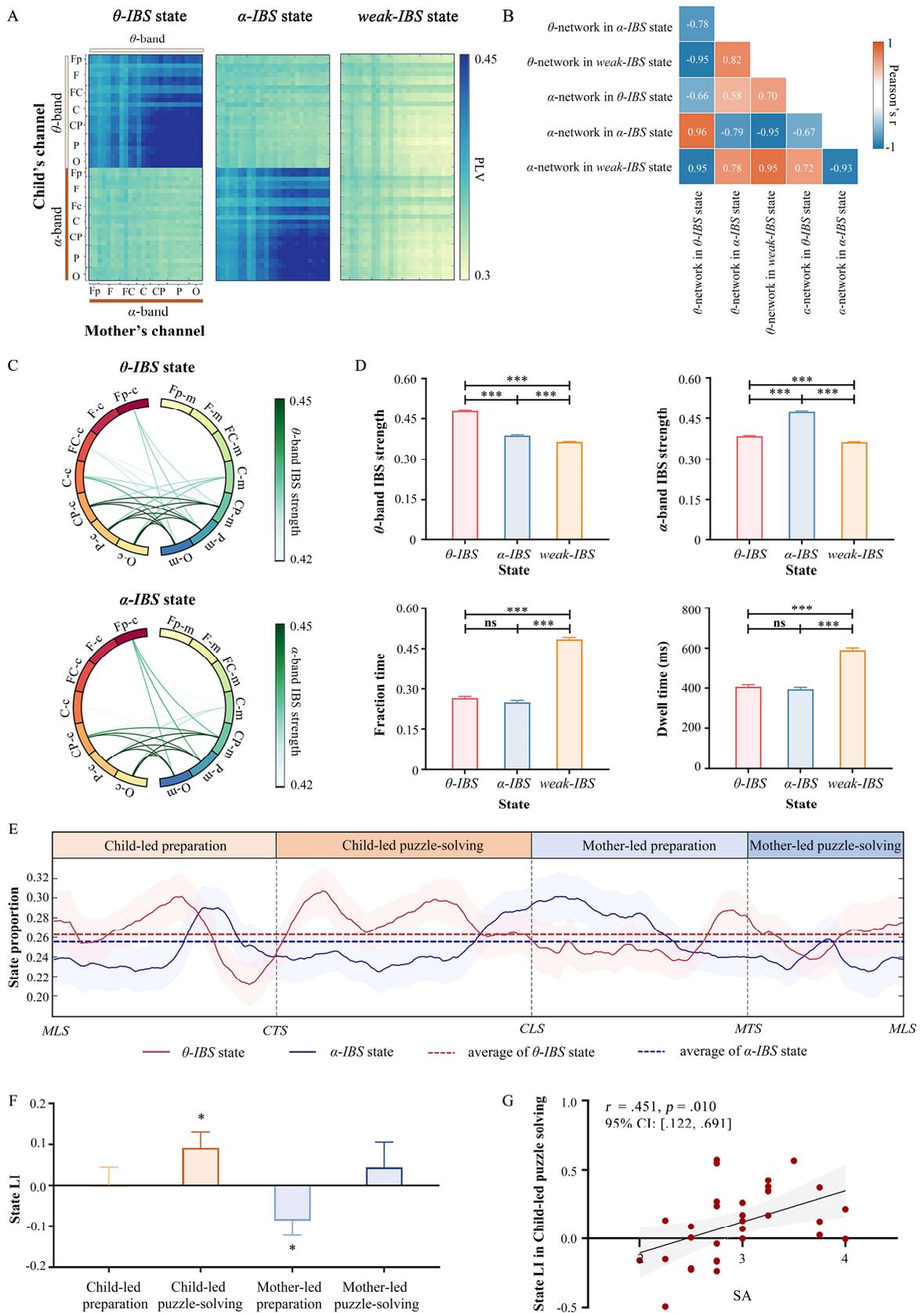

**Figure 2. Three dynamic states of inter-brain synchrony (IBS) during mother-child cooperation (n = 40).** (A) An overview of three dynamic states. The first state is characterized

by a high IBS at θ band (*θ-IBS* state), the second state is characterized by a high IBS at α band (*α-IBS* state), and the third one is characterized by overwhelming weak IBS at both θ and α bands (weak-IBS state). (B) Spatial similarity between inter-brain subnetworks of different IBS states. (C) Average strength of inter-brain connections across different brain regions in θ-IBS and α-IBS states. Fp, prefrontal; F, frontal; FC, frontocentral; C, central; CP, centroparietal; P, parietal; O, occipital; c, child; m, mother. (D) shows the mean strength of IBS, fraction time, and dwell time of each state. ***$p_{Bonf}$ < .001. (E) Group-averaged proportion of θ-IBS state and α-IBS state during the turn-taking round. Only the mother-child dyads with more than six turn-taking rounds (n = 32) were included. A round includes child-led preparation, child-led puzzle-solving, mother-led preparation, and mother-led puzzle-solving phases, which are divided by the events of Mother's finger Lifting off the Screen (MLS), Child's finger Touching the Screen (CTS), Child's finger Lifting off the Screen (CLS), and Mother's finger Touching the Screen (MTS). Shaded areas show the standard error of the means. (F) The laterality index (LI) between the faction time of θ-IBS state and α-IBS state. The LI in each phase is compared with the LI across the whole turn-taking round (whole-round LI) using paired t-test with false discovery rate (FDR) correction, and the changes in LI relative to whole-round LI are shown in the panel. *$p_{FDR}$ < .05 (G) The significant positive correlation between the LI in the child-led puzzle-solving phase and the score of secure attachment (SA).

**Changes of EEG power along with IBS state transitions**

It can be noted that the among-state difference was especially evident between children's and mothers' central, parietal, and occipital electrodes. Therefore, EEG signals in these electrodes were selected for EEG spectral analysis, and changes in EEG power along with IBS state transitions were measured using the time-lagged cross-correlation. A positive (negative) correlation with a specific IBS state indicated an increase (decrease) of EEG power when switching to the IBS state and a decrease (increase) of EEG power when switching from the IBS state, with the time-lag (*t*) on the x-axis **(Figure 3)**.

The results showed that *θ-IBS* state **(Figure 3A-C)** was positively correlated with children's EEG power at 3-4 Hz (with a peak correlation at *t* = 11.55 ± 14.8 ms) and mothers' EEG power at 3-4 Hz (with a peak correlation at *t* = -11.8 ± 13.2 ms), and negatively correlated with mothers' EEG power at 6-7 Hz (with a peak correlation at *t* = -6.0 ± 15.1 ms); while *α-IBS* state **(Figure 3D-F)** was positively correlated with children's EEG power at 6-7 Hz (with a peak correlation at *t* = -15.5 ± 15.9 ms) and mothers' EEG power at 6-8 Hz (with a peak correlation at *t* = -18.8

± 14.1 ms). All these correlations reached a peak value around $t = 0$ (all $p > .156$ compared with $t = 0$ using the Wilcoxon rank-sum test), suggesting that the changes in EEG power and IBS state were generally synchronous. The cluster-based permutation test showed that all these results differed significantly from chance (all $p < .001$, |Cohen's d| $> 18.61$).

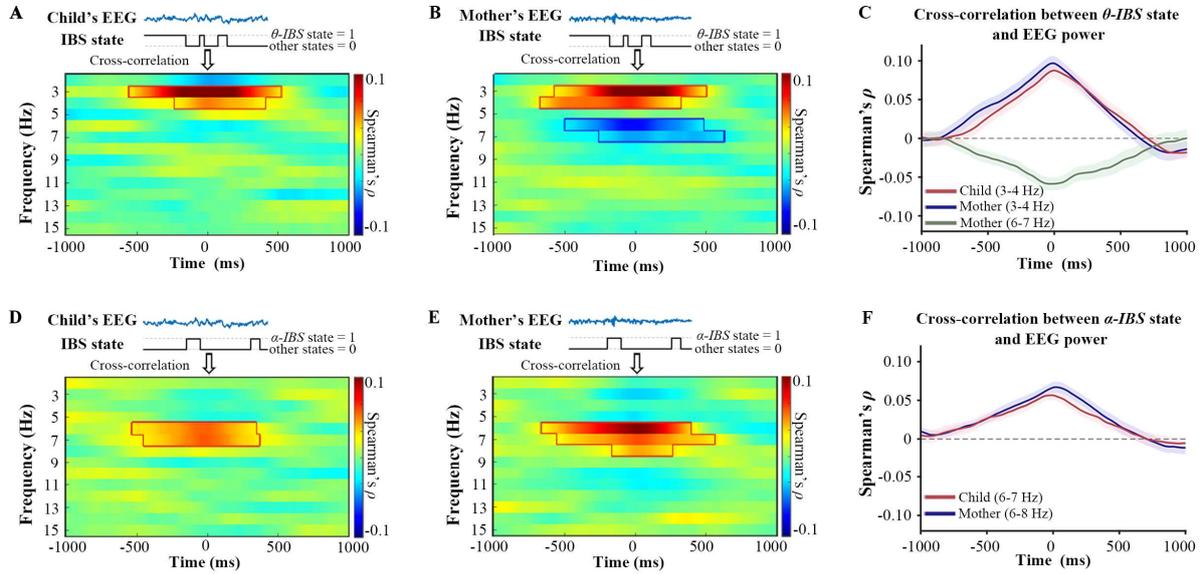

**Figure 3. Time-lagged cross-correlations between EEG power and IBS states during mother-child cooperation (n = 40).** The cross-correlations are evaluated using Spearman's $\rho$ coefficient. The time lag between EEG power and IBS state is shown on the x-axis. The time-frequency windows with cross-correlation that differ significantly from chance from the cluster-based permutation test are marked with a red border. (A) The cross-correlation between child's EEG power and $\theta$-IBS state. (B) The cross-correlation between mother's EEG power and $\theta$-IBS state. (C) The cross-correlation with $\theta$-IBS state for those frequency bands identified from the cluster-based permutation test (child: 3-4 Hz; mother: 3-4 Hz and 6-7 Hz). (D) The cross-correlation between child's EEG power and $\alpha$-IBS state. (E) The cross-correlation between mother's EEG power and $\alpha$-IBS state. (F) The cross-correlation with $\alpha$-IBS state for those frequency bands identified from the cluster-based permutation test (child: 6-7 Hz; mother: 6-8 Hz).

**Dynamic characteristics of IBS in different groups and interactive conditions**

Three similar dynamic states were clustered when combining the data from the mother-child dyads (n = 18) and the stranger-child dyads (n = 19) in both individual and cooperative conditions **(Figure S1)**. Significant between-condition and between-group differences in dynamic characteristics were explored by two-way mixed ANOVA of each dynamic measure of each state **(Figure 4)**, taking *group* (two levels: mother-child dyads and stranger-child dyads) as the between-subject factor, and *condition* (two levels: individual and cooperative) as the within-subject factor.

The between-group differences were mainly found in the strength of IBS. The IBS strength of each state was significantly stronger in the mother-child dyads compared with the stranger-child dyads (main effect of *group*: all $p < .03$, $\eta_p^2 > 0.137$), except the $\alpha$-band IBS strength of $\alpha$-*IBS* state (main effect of *group*: $p = .087$, $\eta_p^2 = 0.081$) and the $\theta$-band IBS strength of *weak-IBS* state (main effect of *group*: $p = .135$, $\eta_p^2 = 0.063$). In addition, a significant main effect of *condition* was observed on the $\theta$-band IBS strength of $\theta$-*IBS* state (F (1,35) = 7.71, $p = .009$, $\eta_p^2 = 0.180$, **Figure 4A**). However, this between-condition difference was only significant in the mother-child dyads (individual condition: 0.468 ± 0.002, cooperative condition: 0.478 ± 0.003, t(17)= -3.225, $p_{Bonf} = .016$) but not in the stranger-child dyads (individual condition: 0.466 ± 0.002, cooperative condition: 0.468 ± 0.003, t(17)= -0.667, $p_{Bonf} = 1.000$), in accordance with a marginally significant *group* × *condition* interaction (F (1,35) = 3.41, $p = .073$, $\eta_p^2 = 0.089$). The $\alpha$-band IBS strength of $\alpha$-*IBS* state also exhibited a significant *group* × *condition* interaction (F (1,35) = 12.37, $p = .001$, $\eta_p^2 = 0.261$, **Figure 4B**). Post-hoc analysis found an enhancement of $\alpha$-band IBS strength from individual to cooperative condition specifically in the mother-child dyads (individual condition: 0.467 ± 0.002, cooperative condition: 0.481 ± 0.004, t(17)= -3.215, $p_{Bonf} = .017$), but not in the stranger-child dyads (individual condition: 0.472 ± 0.002, cooperative condition: 0.465 ± 0.004, t(17) = 1.739, $p_{Bonf} = .545$).

The between-condition differences were also found in the fraction time (the main effect of *condition*: F (1,35) = 6.849, $p = .013$, $\eta_p^2 = 0.164$, **Figure 4C**) and the dwell time (the main effect of *condition*: F (1,35) = 9.28, $p = .004$, $\eta_p^2 = 0.210$, **Figure 4D**) of $\theta$-*IBS* state. The fraction time and the dwell time of $\theta$-*IBS* state were significantly longer in cooperation (fraction time: 0.289 ± 0.005, dwell time: 424.94 ± 8.99 ms) than those in the individual condition (fraction time: 0.269 ± 0.006, dwell time: 389.25 ± 6.85 ms). These between-condition differences were comparable in two groups (*group* × *condition* interaction: both $p > .8$, $\eta_p^2 < 0.002$). Moreover, the fraction time and the dwell time of $\theta$-*IBS* state during mother-child cooperation were related to the mother-child relationship in daily life. Specifically, the fraction time of $\theta$-*IBS* state was found to be significantly and positively correlated with the score of the Chinese Parent-Child Interaction Scale (CPCIS) (Pearson's $r = .601$, 95% CI: [.186, .834], $p$

= .008, **Figure 4E**). The dwell time of *θ-IBS* state was significantly and positively correlated with the SA score (Pearson's *r* = .474, 95% CI: [.009, .770], *p* = .047, **Figure 4F**).

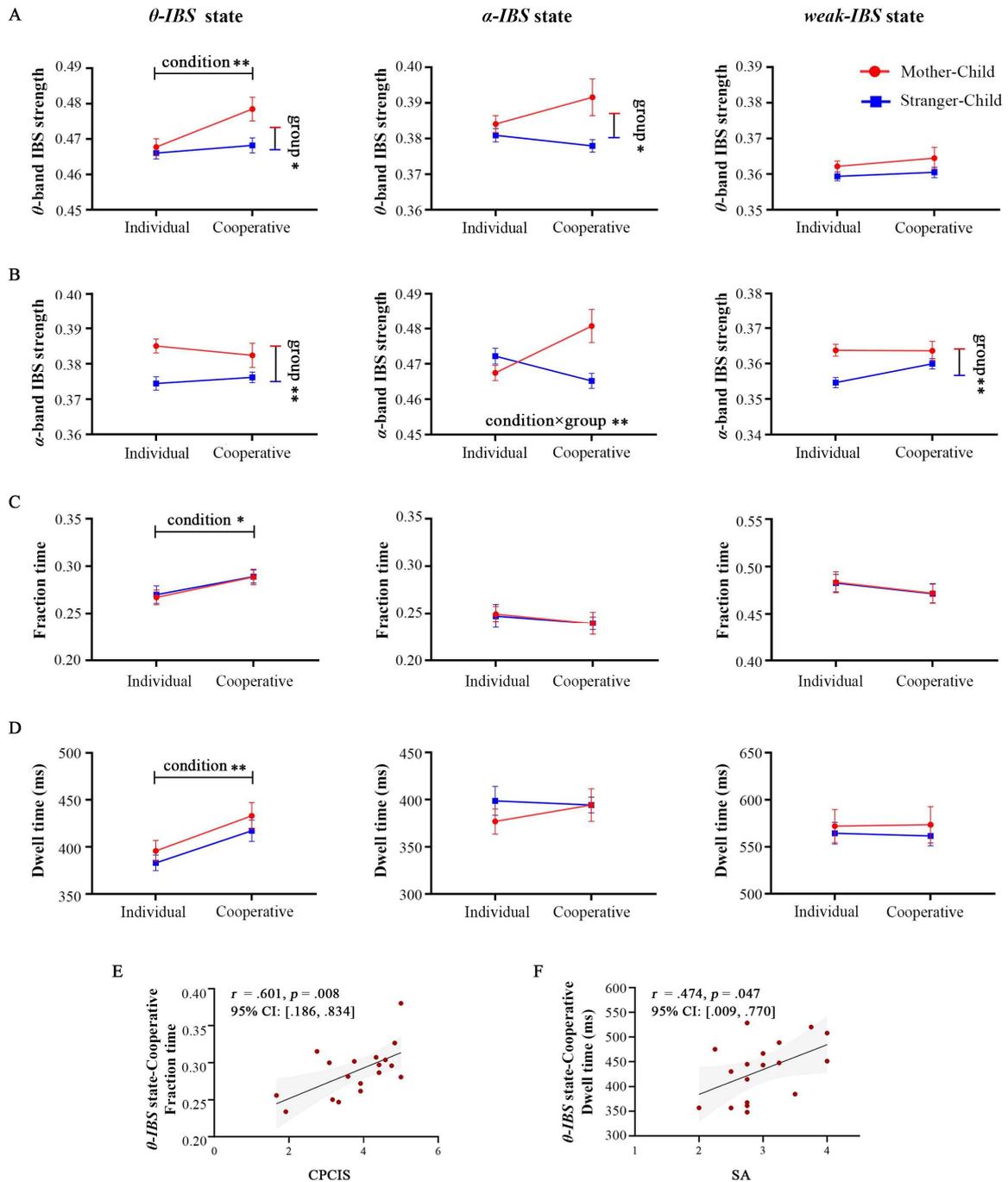

**Figure 4. Comparison of characteristics of each dynamic state between the mother-child dyads (n = 18) and the stranger-child dyads (n = 19) and between the individual and cooperative conditions.** (A) The mean strength of IBS at *θ* band. (B) The mean strength of IBS at *α* band. (C) Fraction time. (D) Dwell time. Significant group and condition main effects, as well as their interaction effects, are marked. **p* < .05, ***p* < .01. (E) The significant positive

correlation between the fraction time of θ-IBS state in mother-child cooperation and the score of Chinese Parent-Child Interaction Scale (CPCIS). (F) The significant positive correlation between the dwell time of θ-IBS state in mother-child cooperation and the score of secure attachment (SA).

## 3. Discussion

This study delicately constructed a dual-band inter-brain network framework that comprised θ-band and α-band subnetworks and, for the first time, explored the dynamic states of IBS during mother-child turn-taking cooperation. This approach overcame the limitations of previous static and single-band analysis in capturing the transient inter-brain connectivity fluctuations within or between frequency bands, and was able to draw a vivid and dynamic picture of mother-child inter-brain interactions that θ and α oscillations synchronize alternately across two brains.

Firstly, the dynamic inter-brain network analysis demonstrated the existence of θ-IBS state and α-IBS state during mother-child interactions (**Hypothesis 1**). The θ rhythm is vital in endogenously driven attention, social brain development, and parent-child interaction of infants and preschoolers (Begus and Bonawitz, 2020). For example, infants' θ power can predict their subsequent attentional behaviors (Wass et al., 2018). The θ networks, but not α networks, were found to experience strong global reconfigurations in the first year of life and exhibit increased sensitivity in differentiating social from non-social stimuli (Van Der Velde et al., 2021). Additionally, the θ connectivity in infants was linked to maternal caregiving quality, such as responsiveness, positive emotional tone, and reciprocity (Perone and Gartstein, 2019). However, during development, the α rhythm becomes increasingly involved in attention, social cognitive functioning, and emotional processing, supplanting the θ rhythm. This transition was evidenced by the finding that the EEG frequencies for neural-to-attention predictiveness were shifted from the θ range in infants to the α range in adults (Wass et al., 2018). From an inter-brain perspective, maternal frontal α asymmetry in the resting state was predictive of the infant frontal asymmetry in emotion-eliciting conditions (Krzeczkowski et al., 2022), and the α-band inter-brain network was influenced by maternal emotions (Santamaria et al., 2020). Therefore, considering the distinct social brain rhythms for children and adults, θ-IBS and α-IBS can be generally interpreted as child-centered and mother-centered IBS states, respectively, in which mothers and children align neural activity towards each other during interactions.

This interpretation of the roles of *θ-IBS* and *α-IBS* in mother-child inter-brain communication is supported by our observations in IBS changes along with child-led and mother-led behaviors in the turn-taking cooperation. Generally, the turn-taking procedure consists of four interactive phases: the child-led preparation and puzzle-solving phases and the mother-led preparation and puzzle-solving phases. First, by calculating the average IBS in different role contexts, our recent study in static analysis found that when mothers were observing their children solve a puzzle (i.e., child-led puzzle-solving phase), the strength of IBS was increased with the degree of interaction specifically in the *θ* band; and when their roles were switched (i.e., mother-led puzzle-solving phase), the interaction-induced IBS strengthening was shifted to the *α* range (Li et al., 2024). Next, in this study, from the dynamic aspect, we found that the LI between the fraction time of *θ-IBS* state and *α-IBS* state dynamically changed along with interactive behaviors. Mother-child IBS showed a lateralization towards *θ-IBS* state when mothers were observing their children solve a puzzle (i.e., child-led puzzle-solving phase). This aligns with the fact that mothers were typically more concerned about their children's performance on the puzzle-solving task. Consequently, mothers attempted to track their children's attention and predict children's intention of puzzle-solving, resulting in a higher proportion of *θ-IBS* state occurring in this phase. Conversely, a lateralization towards *α-IBS* state was found when children were observing their mothers move the right hand to pick up a puzzle piece in preparation for puzzle-solving (i.e., mother-led preparation phase). One possible explanation is that children were likely to care more about which color or shape of the puzzle piece their mothers chose, as evidenced by children's occasional whispers like "Mum, you take the puzzle piece that I want" during the interactions. The association between IBS states and behaviors well substantiates **Hypothesis 2**.

The cross-correlation between EEG power and IBS states suggested an increase of EEG power in the corresponding frequencies, indicating EEG frequency-shifts to those frequencies, along with IBS state transitions (**Hypothesis 3**). The EEG frequency-shift to *θ* band in both children and mothers could be interpreted as a prerequisite for mother-toward-child neural alignment in order to achieve *θ-IBS* state, which further supports our inference of *θ*-IBS as a child-centered interactive state. Wass et al. reported a similar down-shift of mothers' EEG frequency during mother-infant joint play, and the mothers' *θ* power significantly tracked infants' attention (Wass et al., 2018). Our study extends this finding by linking it to mother-toward-child neural alignment. Combining Wass et al.'s findings with ours, we speculate that in order to track children's attention and predict children's intentions during interactions, mothers needed to establish a neural alignment with their children by shifting their EEG frequencies to children's θ range, which was manifested by an achievement of θ-IBS state **(Figure 5A)**. Similarly, the EEG frequency-shift to *α* band in dyads was inferred to be necessary for child-toward-mother neural alignment in the mother-centered *α-IBS* state, potentially linked to children's efforts to track mothers' attention and anticipate mothers' intentions **(Figure 5B)**. This capability likely reflects the children's social learning ability, demonstrating an early development of skills for understanding and responding to social cues. This interpretation is evidenced by the positive correlation found between mother-infant *α-IBS* and the social learning of infants (Leong et al., 2019). Overall, the EEG frequency-shifts accompanying IBS state transitions, to a great extent, enhance our understanding of how IBS is established and dynamically changed during mother-child interactions.

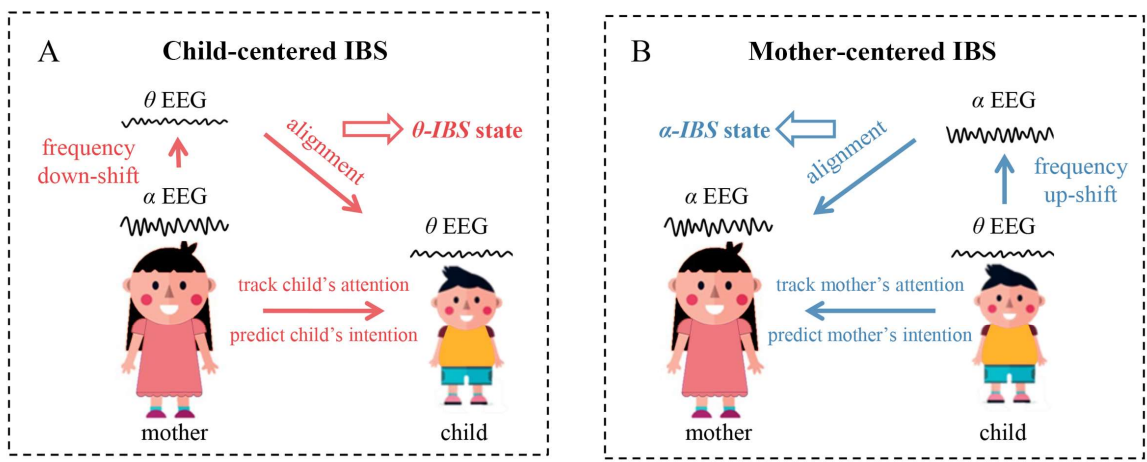

**Figure 5. The proposed model for *θ-IBS* (A) and *α-IBS* (B) in mother-child interactions.** (A) When mothers attempt to track their children's attention and/or predict their intentions, they will down-shift their EEG frequencies to align with their children's *θ* oscillations, achieving a child-centered *θ-IBS* state. (B) When children endeavor to track or anticipate the mothers' attention or intentions, they will up-shift their EEG frequencies to align with their mothers' EEG, achieving a mother-centered *α-IBS* state.

In addition to *θ-IBS* and *α-IBS* states, we also identified a weakly-connected state in which IBS was significantly lower in either *θ* or *α* bands. It was not shown that two brains are always highly synchronized during the process of mother-child interactions. Instead, they were weakly inter-connected for approximately half of the time. A similar sparse inter-brain state was observed in a dynamic study on creative and non-creative communications of adults using fNIRS hyperscanning (Wang et al., 2022). The existence of the weak-IBS state might imply the difficulty for human brains in maintaining a long-term and efficient connective state during interactions. Mechanistically, the shift between strong-IBS and weak-IBS states might be associated with attention switching between external stimuli and internal thoughts during interactions, which has been confirmed in fMRI studies (Kucyi et al., 2020, 2016).

It should be noted that the three IBS states were not exclusively observed in mother-child cooperation, but were evident in both mother-child and stranger-child dyads across both individual and cooperative conditions. Nonetheless, in any state, the mother-child dyads showed stronger IBS compared with the stranger-child dyads, consolidating the conclusions from previous static studies (Endevelt-Shapira et al., 2021; Reindl et al., 2018; Zhao et al., 2021). Besides strength, other dynamic metrics (i.e., fraction time and dwell time) showed no significant difference between the mother-child and stranger-child dyads. However, longer fraction time and dwell time of $\theta$-IBS state, indicating more frequent and stable IBS at $\theta$ band, were observed in cooperation than the individual condition. Moreover, the fraction time and the dwell time of $\theta$-IBS state in mother-child cooperation also correlated with the CPCIS score and SA score. Especially, the degree of lateralization to $\theta$-IBS state during the child-led interactions was positively correlated with the SA score. These results not only endorse the important role of mother-to-child neural alignment in their interactions, but also demonstrate that mothers with stronger emotional bonds and more frequent interactions with their children were more capable of shifting their EEG frequencies to their children's $\theta$ band, thereby achieving neural alignment with their children. This interpretation is also echoed by previous behavioral studies that connected mother-child attachment with maternal responsiveness. Mothers with secure attachment to their children were more responsive to children's cues; while resistant attachment style contributed to poorly coordinated interaction in which mothers were minimally involved, unresponsive, or intrusive (Crandell et al., 1997; Isabella and Belsky, 1991). Our findings thus link the mother-child behavioral connection with their neural alignment, illustrating a possible neural mechanism for how the mother-child relationship affects responsive interactions.

There remain some limitations in our study. This turn-taking experimental paradigm does not fully capture the natural cooperation seen in daily life, and participants were instructed to minimize talking, physical touching, and interactive motions to reduce EEG artifacts. Additionally, the tangram puzzle-solving game was relatively easy for most preschoolers. Therefore, it remains unknown whether other IBS states are involved in more natural and reciprocal interactive behaviors, such as when children seek assistance from their mothers when facing difficulties. Further dynamic studies are thus encouraged in more natural and reciprocal interaction settings.

## 4. Methods

**Datasets**

This study was conducted based on our two datasets. One was from our recent study (Li et al., 2024), which included dual-EEG data of both mother-child and stranger-child dyads in both cooperative and individual conditions. The other only included dual-EEG data in cooperation. Participant dyads in both datasets participated in the same tangram puzzle-solving video game on a touch screen (23.8-inch, NEWTAP Inc., China), and dual-EEG signals were recorded using the same equipment and parameters, i.e., two synchronized 32-channel mobile EEG systems (NeuSen.W32, Neuracle, China), at a sample rate of 1000 Hz, with FCz as the reference and AFz as the ground. In the cooperative condition, two partners of a dyad were seated next to each other and solved four tangram templates jointly by moving the puzzle pieces by turn; while in the individual condition, one participant solved all the puzzle pieces in the template alone (two templates for each participant) and his/her partner was seated in another room watching the screen recording from the actor. Each tangram template consisted of seven pieces of geometric shape puzzles. The participants were asked to move puzzles using their right hand steadily and keep attending and silent during the task. The details of experiment design and settings were described in our previous study (Li et al., 2024). Both experiments were approved by the Review Board of Ethics Committee of Shanghai Children's Medical Center (SCMCIRB-K2020108-1 and SCMCIRB-K2023099-1).

To test three hypotheses, we selected dual-EEG data of mother-child dyads in cooperation from both datasets. Totally, we selected 40 dyads (children: children: 21 boys and 19 girls, mean age: $3.76 \pm 0.06$ years) according to the following criteria: (1) children's ages should be between 3-4 years; (2) both mother and child in the dyad should be right-handed according to the Edinburgh handedness inventory, with normal or corrected-to-normal vision, and had no known history of neurological or psychiatric disorders; (3) child in the dyad classified as secure attachment according to the child attachment scale (Ye, 2011) and with normal psychosocial well-being according to the Chinese version of the Strengths and Difficulties Questionnaire (SDQ) (Goodman and Goodman, 2009); (4) dyads followed the instruction to move puzzles using their right hand across the whole task.

To explore the dynamic characteristics in different groups and interactive conditions, we used the first dual-EEG dataset from 18 mother-child dyads (children: 10 boys and 8 girls, mean age: 3.83 ± 0.13 years) and 19 stranger-child dyads (children: 6 boys and 13 girls, mean age: 3.88 ± 0.11 years) in both cooperative and individual conditions, which also met the inclusion criteria.

**Mother-child relationship measures**

For each mother-child dyad, the mother-child relationship was assessed with three measures, i.e., child-mother attachment, parent-child dysfunctional interaction (P-CDI), and CPCIS. All these questionnaires were self-reported by the mother.

The child-mother attachment mainly reflects the child's attachment to the mother based on the mother's responses about how the child usually interacts with her (Ye, 2011). The questionnaire covers three dimensions of SA, avoidant/disorganized attachment, and ambivalent attachment, scored separately. Child-parent attachment style was assigned to the dimension with the highest score. Due to very few cases of avoidant/disorganized attachment and ambivalent attachment (n = 3) in our datasets, we only included children of SA style in this study, and their SA scores were used for analysis.

P-CDI is a subscale in the Parenting Stress Index-short Form (Yeh et al., 2001). It indicates parents' dissatisfaction with interactions with their children and the degree to which parents find their children unacceptable. The higher the P-CDI score, the more dissatisfaction the mother feels.

CPCIS is a parent-child interaction behavior scale designed to combine the characteristics of Chinese parenting environment (Ip et al., 2018). It measures the averaged frequencies of parent-child interaction in the past month from four dimensions, including learning-related activities, reading activities, recreation activities, and interaction with the environment.

**EEG preprocessing**

The EEG preprocessing was performed using the EEGLAB toolbox in MATLAB (MATLAB R2018b, The MathWorks Inc., Natick, MA) (Delorme and Makeig, 2004). EEG signals were firstly band-pass filtered into 1-30 Hz and notch filtered at 50 Hz. Peripheral temporal channels (T7, T8, Tp9, Tp10) were then excluded from the analysis due to the muscular artifacts, and as a result, a total of 26 channels (Fp1, Fp2, F3, F4, Fz, F7, F8, FC1, FC2, FC5, FC6, C3, C4, Cz,

CP1, CP2, CP5, CP6, P3, P4, Pz, P7, P8, O1, O2, Oz) were left for analysis. Independent component analysis (ICA) was further utilized to remove ocular and muscular artifacts. The epochs with amplitude values exceeding ±100 μV, or when the participants were speaking, moving head intensely, or not attentive, were excluded. Then, the EEG data were filtered into children's $\theta$ (3-6 Hz) and $\alpha$ (6-9 Hz) bands, respectively. Finally, high-quality EEG epochs that consisted of solving more than five consecutive pieces of puzzles were segmented for dynamic analysis. On average, $2.74 \pm 0.14$ epochs with a mean duration of $28.35 \pm 1.18$ s per epoch were segmented for each dyad and each interactive condition.

**Dynamic IBS states construction and measurement**

Dynamic IBS states were estimated utilizing sliding window and $k$-means clustering (Li et al., 2021; Wang et al., 2022). The whole framework is illustrated in Fig.1. First, dual-EEG epochs were extracted using a sliding window with a window size of 1 s and a step size of 0.01 s. Second, in each time window, two inter-brain subnetworks were constructed in $\theta$ and $\alpha$ bands, respectively, by using phase locking value (PLV) (Lachaux et al., 1999), which is commonly used to measure IBS. In this procedure, PLVs for all possible pairs between the child's EEG channels and the adult's EEG channels were calculated, forming a $26 \times 26$ matrix for each frequency band in each time window. Then, $\theta$ and $\alpha$ inter-brain subnetworks were concatenated into a $52 \times 26$ dual-band inter-brain network. Finally, dual-band inter-brain networks from all dyads and all the time windows were clustered using $k$-means clustering. Manhattan distance (L1 distance) function, which has been confirmed effective in high-dimensional neuroimaging data (Aggarwal et al., 2001), was applied to calculate the similarity between clusters. The number of clusters $k$ was determined by the elbow criterion of cluster cost (Allen et al., 2014; Fang et al., 2020), which is the ratio between the within-cluster and between-cluster distance. The clustering was repeated 1000 times with different initializations to escape the local minima, resulting in optimized cluster centroids along with the final states.

The spatial similarity between different IBS states was assessed using Pearson's correlation. For each IBS state, its fraction time, dwell time, and strength of synchrony were measured. The fraction time is defined as the percentage of a given state in total time windows. The dwell time refers to the average time of a state without switching to another one. The strength of synchrony was measured in $\theta$ and $\alpha$ bands, respectively, which was calculated as the mean intensity of the top 20% values in the averaged matrix in each state.

**Dynamic IBS states in different phases of the turn-taking cooperation**

In addition to calculating the overall occurrence of all derived IBS states throughout the entire turn-taking cooperation, we examined their occurrence changes during this procedure. Generally, each turn-taking round can be divided into four phases based on four specific behavior events (i.e., the child's/mother's finger touching/lifting off the screen). The first is the child-led preparation phase, during which the child moves his/her right hand to pick up a puzzle piece in preparation for puzzle-solving. This phase begins after the mother completes the previous puzzle task (indicated by the event of the mother's finger lifting off the screen), and ends when the child's finger touches the screen. Next, the child-led puzzle-solving phase begins, and it ends when the child's finger lifts off the screen. Then, taking turns, the mother-led preparation and puzzle-solving phases occur successively. Once the mother completes her puzzle-solving, the round ends and the next begins.

To analyze the occurrence of *θ-IBS* state and *α-IBS* state in different phases, the IBS state series was firstly segmented into turn-taking cooperation rounds according to the events of the mother's finger lifting off the screen. Here, only dyads with more than six rounds (n = 32) were included in this analysis. Then, the IBS state series in each round was linearly time-warped to align the time points of four behavior events, with these events occurring at the same average latencies. The proportion of different IBS states was calculated for each time point from the time-aligned state series, obtaining a proportion series for each state and each dyad. These proportion series in each state were then averaged across dyads to obtain a group mean (Fig.2E).

We also calculated the LI between the occurrence of *θ-IBS* state and *α-IBS* state, which is defined as

$$LI = \frac{F_{\theta-IBS} - F_{\alpha-IBS}}{F_{\theta-IBS} + F_{\alpha-IBS}}, \qquad (1)$$

where $F_{\theta-IBS}$ and $F_{\alpha-IBS}$ represent the fraction time of *θ-IBS* state and *α-IBS* state, respectively. The LI was calculated for each phase as well as the entire turn-taking round (whole-round LI).

**Time-lag cross-correlation analysis between EEG power and IBS states**

First, we coded different IBS states and calculated EEG power for cross-correlation analysis. Specifically, when calculating the correlation with *θ-IBS* state, *θ-IBS* state was coded as 1 and other states were coded as 0; and when calculating the correlation with *α-IBS* state, *α-IBS* state was coded as 1 and other states were coded as 0. The spectral power of EEG signals in each sliding window was estimated using the fast Fourier transform algorithm, with a frequency resolution of 1 Hz. Then, time-lag cross-correlations between these two continuous and time-synchronized data (i.e., EEG power and IBS-state codes) were calculated using a nonparametric (Spearman's) correlation, from -500 to 500 ms in lags of 10 ms. For each time × frequency cell, individual cross-correlation values were averaged across participants to obtain a group mean.

**Statistical analysis**

After different IBS states had been detected during mother-child cooperation, their dynamic characteristics were compared using one-way repeated ANOVA. Then, these dynamic characteristics were further compared between mother-child and stranger-child dyads and between individual and cooperative conditions, using two-way mixed ANOVAs, taking group and condition as the between-subject and within-subject factors, respectively. Bonferroni correction was applied to correct multiple comparisons in post-hoc analysis. In addition, Pearson's $r$ (or Spearman's $\rho$ if the variables are non-normally distributed) was applied to explore the relationships between the dynamic characteristics of IBS and mother-child relationship measures. The normality of variables was tested using Shapiro-Wilk's test. The LI in each phase was compared with the whole-round LI using a paired t-test with false discovery rate (FDR) correction. Statistical significance was accepted for $p < .05$. All data were presented as mean ± standard error of the mean.

A bootstrapping method with a cluster-based correction was applied in the statistical analysis of cross-correlation, to identify the cross-correlation between EEG power and IBS states that is significantly different by chance. We calculated the spurious cross-correlation by randomly phase-shuffling the EEG power series for 1000 times, generating 1000 time × frequency distributions of spurious cross-correlation. For each time × frequency cell, a real or spurious correlation was compared against the distribution of 1000 spurious correlations to test the significance ($p < .01$). Contiguously significant cells were clustered in time-frequency space, and the size of significant clusters was identified for each iteration. The size of the largest cluster obtained from each shuffling constituted a distribution of 1000 cluster-sizes by bootstrapping. Finally, the cluster size obtained from real signals was compared against the distribution of 1000 cluster-sizes by bootstrapping, and those with sizes significantly larger than chance were determined ($p < .01$). For the EEG frequency bands with significant correlation with IBS states, the time-lag for peak correlation was further detected for each participant and statistically compared with $t = 0$ using Wilcoxon rank-sum test.

**Acknowledgements**

We thank Dr. Xiaoning Sun and Dr. Jin Zhao in Shanghai Children's Medical Center for their valuable advice on assessment of parent-child interaction. We also thank Jiawen Zhang, Dan Wang, Jiaqi Wang, Ziping Xing, Xintong Chen, Yufeng Liu, Xiaoyan Zhang and Zihan Yang for the help in data collection, and Ira Marriott Haresign for the help in statistics.

**Fundings**

This work was supported by the National Natural Science Foundation of China (No. 62371285), Fundamental Research Funds for the Central Universities (No. YG2023QNB21), and the Innovative Research Team of High-level Local Universities in Shanghai.

# Supplementary Information

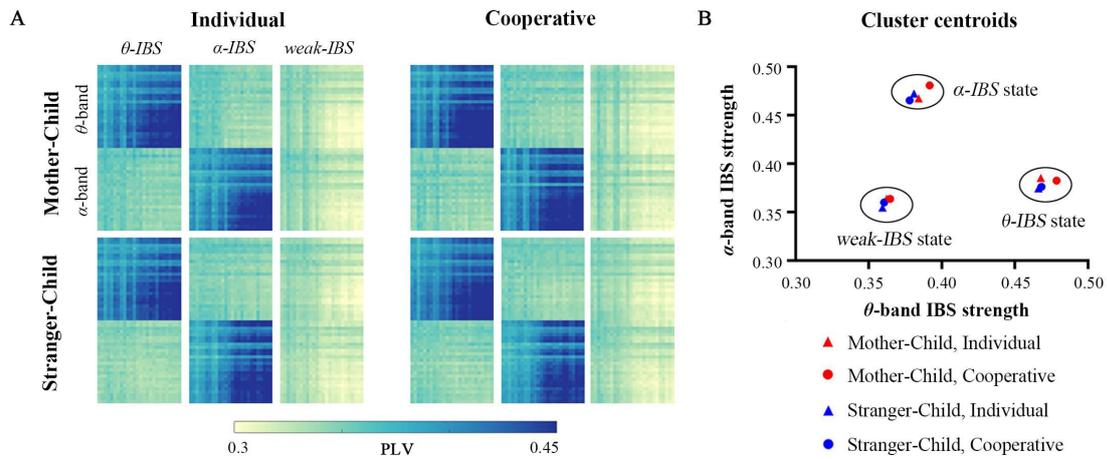

**Figure S1.** Dynamic IBS states in the mother-child dyads (n = 18) and the stranger-child dyads (n = 19), and in their individual and cooperative conditions. (A) An overview of three dynamic states in each group and each condition. (B) Cluster centroids of each dynamic state in each group and each condition.